\documentclass[]{aa}
\usepackage{amsmath}
\usepackage{graphicx}
\usepackage{txfonts}

\begin{document}

   \title{Dynamical evolution of escaped plutinos, another source of Centaurs.}

   \author{R. P. Di Sisto
           \thanks{romina@fcaglp.unlp.edu.ar}
           A. Brunini
          \and
           G. C. de El\'\i a 
          }

   \offprints{R. P. Di Sisto
    }

  \institute{Facultad de Ciencias Astron\'omicas y Geof\'\i sicas, Universidad 
Nacional de La Plata and \\
Instituto de Astrof\'{\i}sica de La Plata, CCT La Plata-CONICET-UNLP \\
   Paseo del Bosque S/N (1900), La Plata, Argentina.
                }

   \date{Received / Accepted}


\abstract
{}
{It was shown in previous works  the  existence of weakly chaotic orbits 
in the plutino population that diffuse very slowly. These orbits correspond 
to long-term plutino escapers and then represent the plutinos that are escaping from 
the resonance at present. In this paper we perform
numerical simulations in order to explore the dynamical evolution of 
plutinos recently escaped from the resonance.  
}
{The numerical simulations were divided in two parts. 
 In the first one we evolved $20,000$ test particles 
in the resonance in order to detect and select the long-term escapers.
In the second one, we numerically integrate the 
selected  escaped plutinos in order to study their dynamical post 
escaped behavior.
}
{Our main results include the characterization of the routes 
of escape of plutinos and  their evolution in the Centaur zone. 
We obtained a present rate of escape of plutinos between 1 and 10 every 10 years.  
 The escaped plutinos have a mean lifetime in the Centaur zone of 
$ 108$ Myr and their contribution to the Centaur population would be a fraction 
of less than  $6 \%$  of the total Centaur population. In this way, escaped plutinos 
would be a secondary source of Centaurs.
}
{}


\authorrunning{R. P. Di Sisto, A. Brunini \&
               G. C. de El\'\i a 
               }
\titlerunning{
Dynamical evolution of escaped plutinos
                             }

\maketitle
\section{Introduction}

In the last few years the number of  observed transneptunian objects has 
enormously grown thanks to the progresses in the astronomical observations. This fact 
has allow to define, in a more rigorous way, the different dynamical classes 
previously identified in the first years of discoveries. 
The transneptunian region (TNR) can be structured into 
4 dynamical classes  (Chiang et al. \cite{Chiang07}).
The Resonant Objects are those in   
mean motion resonance with Neptune, the Classical Objects are those non-resonant 
objects with semimajor axis $a$ greater than $\sim$42 AU and low eccentricity orbits, the 
Scattered Disk Objects (SDOs) with perihelion distances $q > 30$ AU and large 
eccentricities, and the Centaurs 
Objects. This last group has perihelion distances inside the orbit of Neptune 
 and they are transitory objects descendants of the other 3 classes, mainly
from the SDOs, recently dislodge from the transneptunian zone by planetary 
perturbations (Levison \& Duncan \cite{Levison97},  Tiscareno \& Malhotra 
\cite{Tiscareno03}), Di Sisto \& Brunini \cite
{Disisto07} ).  Centaurs are sometimes defined according to 
their aphelion distance or to their semimajor axis ($a$), as for example the 
nomenclature of Gladman et al  \cite{Gladman08} that uses $a <a_N$ (where $a_N$ 
is Neptune's semimajor axis),  being then objects entirely in the giant planet 
zone. Nevertheless,  it is generally accepted that they are objects
which enter the planetary region from the TNR,
 evolve in the giant planetary zone and a fraction of them enter 
the zone interior to the Jupiter's orbit becoming Jupiter Family Comets (JFCs).

The resonant transneptunian population most densely populated  are 
the plutinos, that are trapped into the 2:3 mean 
motion resonance with Neptune, being Pluto its most representative member. 
Some of the plutinos cross Neptune's orbit and hence  one 
might think  they would be sensitive  
to strong perturbations during close encounters with that planet. 
However conjunctions occur near plutinos's aphelions 
and then close encounters do not occur. This fact provides a stable 
configuration of the resonance, with the critical angle 
$\sigma = 3 \lambda - 2 \lambda_N - \varpi$ librating around $180^{\circ}$, 
where $\lambda$ and $\varpi$ are the mean longitude and the 
longitude of perihelion of the plutino, and $\lambda_N$ is the mean longitude 
of Neptune.

Duncan et al. (\cite{Duncan95}) analyze  the  dynamical structure of the transneptunian 
region through numerical simulations.  They numerically integrate
 thousands of  test transneptunian bodies in order to  study their 
dynamical behavior and to determine which regions may be
potential sources of the Jupiter family comets that we see today. With
regards to  the Plutinos, Duncan et al. (\cite{Duncan95}) showed that the boundaries 
of this long-lived mean motion resonance have a time scale for 
 instability of the order of the  age of the Solar System, and so 
the long-term erosion of the particles is rather gradual and therefore 
must be continuing at the current epoch. The existence of these currently 
unstable orbits  may be related to the origin of the observed Jupiter family comets.

Morbidelli (\cite{Morby97}) studied the dynamical structure of the 2:3 resonance with 
Neptune in order to analyze possible diffusive phenomena and their relation to the 
existence of long-term escape trajectories. 
 He performed numerical simulations integrating the 
evolution of 150 test particles initially covering the 2:3 resonance with   
eccentricities up to $e = 0.3$ and inclinations less than $i = 5 ^ {\circ }$, for 4 byr.
The  author  found regular orbits that never escaped from the resonance 
present only at moderate eccentricity and small  amplitude of
libration,  and  a  strong   chaotic  region  at  large  amplitude  of
libration, which is quickly depleted and seems to be generated for the
interaction  between the  $\nu_{18}$ and  kozai resonances. Moreover,
Morbidelli (\cite{Morby97}) showed the existence of weakly chaotic orbits 
that diffuse very slowly and finally dive into the strong chaotic region. 
They are then long-term escapers and then plutinos recently escaped 
from the resonance. The source zone  of those particles   produce Neptune-
encountering bodies at the current epoch of the Solar System and 
should be an active source of Centaurs and comets at present. He found 
that 10 $\%$ of the bodies initially in that  weakly chaotic zone are delivered 
to Neptune in the last $10^9$ years. 

The first  attempt at including  the gravitational influence  of Pluto
into numerical models of the  dynamics of plutinos was developed by Yu
\& Tremaine (\cite{Yu99}). These  authors suggested that the effect of
Pluto,  ejecting objects  from the  2:3 resonance  to Neptune-crossing
orbits, may contribute to or  even dominate the flux of JFCs.
  At the  same  time, Nesvorn\'y  et al. (\cite{Nesvorny00}) analyzed the effect of 
Pluto on the 2:3 resonant orbits. They found that  Pluto produces a 
large excitation of the libration amplitudes  in the 2:3 resonance. 
However they estimated that the flux rate that contributes to the flux 
of short period comets is about 1 $\%$ of the 2:3 resonant population 
per $10 ^ {8 }$ years  which is about the same value 
as the flux obtained by Morbidelli  (\cite{Morby97}) without Pluto. 

In the early 2000s,  Melita \& Brunini (\cite{Melita00}) developed a
comparative  study of  mean motion  resonances in  the transneptunian
region. These  authors used the  frequency-map-analysis method (Laskar
\cite{Laskar93}) in  order to describe the dynamical  structure of the
2:3,  3:5 and  1:2  Neptune resonances,  at  39.5, 42.3  and 47.7  AU,
respectively.  Particularly,  Melita  \&  Brunini (\cite{Melita00})
showed  that the  2:3 resonance  presents  a very  robust stable  zone
primarily at  low inclinations, where  a great number of  the observed
plutinos are distributed. They suggested that the existence
of  plutinos in  very unstable  regions can  be explained  by physical
collisions  or  gravitational   encounters  with  other  plutinos.

de El\'{\i}a et al. (\cite{deelia08}) performed a collisional evolution of 
plutinos and obtained a plutino removal by their collisional evolution 
of 2 plutinos with $R>1$ km every 10000 years or a flux rate of escape of 
0.5 $\%$ of plutinos in $10^{10}$ years.

Very recently, Tiscareno \& Malhotra (\cite{Tiscareno09}) carried out 1-Gyr 
numerical integrations to study the characteristics of the 2:3 and 1:2 mean motion 
resonances with Neptune. Their main results include maps of resonance stability 
for a whole range of eccentricities and inclinations. 
 They made integrations with and without Pluto, and concluded that it has only 
modest effects on the Plutino population. They calculated the fraction 
of remaining plutinos in the resonance as a function of time and extrapolated this 
fraction after 4 Gyr and also evaluated the fate of escaped particles.

From those previous works, plutino removal by ``dynamic'' is much greater 
than plutino removal by collisions.   
We refer to ``dynamical'',  those numerical simulations that take into account 
the gravitational forces to follow up the evolution of a particle and occasionally 
cause the removal. Then in this work, we perform ``dynamical'' numerical simulations
in order to describe and characterize the routes of 
escape of plutinos and their contribution to the other minor bodies populations
 of the Solar System, specially to Centaurs.

\section{The numerical runs}

The goal of our work is to characterize the post escaped evolution 
of plutinos and the presence of plutinos in other small body population 
in the current Solar System. So we need to identify the plutinos that have
recently escaped from the resonance. Morbidelli (\cite{Morby97}) studied
the dynamical structure of the 2:3 mean motion resonance  with Neptune in order 
to analyze possible diffusive phenomena and their relation to the 
existence of long-term escape trajectories from the 2:3 resonance. 
The  author  showed  the  existence  of weakly chaotic orbits 
that diffuse very slowly finally diving into a strong chaotic region.  
These orbits corresponds then to long-term escapers i.e. plutinos recently escaped 
from the resonance. Then  we divided the numerical simulations in two parts. 
At first we develop a numerical simulation 
of plutinos in the resonance in order to detect those plutinos that  have 
recently escaped from the resonance. Second we perform a numerical simulation of the 
selected escaped plutinos in order to study their dynamical post escape behavior.
In the following subsections we will describe both integrations.

\subsection{Pre-runs. The integration in the resonance}

In order to detect the long-term escapers from the plutino population, we performed 
a numerical integration following the study of Morbidelli (\cite{Morby97}). 
We integrate the evolution of $20,000$ test particles under the gravitational influence of 
the Sun and the four giant planets over 4.5 Gyr with an integration step of 0.5 years
using the hybrid integrator EVORB (Fern\'andez et al. \cite{Fernandez02}).

 We  set the initial orbital elements such that they cover the 
present observational range of orbital elements of plutinos. 
The initial semimajor axis of the particles was set equal to the exact value of the 
resonance $a_i = 39.5$  AU. 
The initial argument of perihelion $\omega$, longitude of node $\Omega$ 
and the mean anomaly $M$ have been chosen at random in the range of $[0^{\circ},360^{\circ}]$ 
in a way that the critical angle $\sigma$ remains between $180^{\circ}$ and 
$330^{\circ}$. Since $\sigma$ librates around $180^{\circ}$, and given the relation 
between $a$ and $\sigma$ in the resonance, the election 
of  $\sigma > 180^{\circ}$ and $a_i = 39.5$ AU.  covers the 2:3 mean motion resonance 
(see Morbidelli \cite{Morby97} for a complete explanation). 
 The maximum limit of $\sigma$ equal to $330^{\circ}$,  was 
selected  taking into account previous papers of Malhotra (\cite{Malhotra96}) , Morbidelli (\cite{Morby97}) 
and Nesvorn\'y and Roig (\cite{NR00}), that state that orbits starting at large 
amplitude of libration ($A_{\sigma}$ ) are in a strong chaotic region, very unstable 
and fast driven to the borders of the resonance. 
In the present paper we are interested in the long term 
escapers from the plutino population and because particles with large initial 
$A_{\sigma}$ will escape at the begining of the integration they will not contribute to 
the long term escaper flux.
 
 The initial eccentricity and inclination of the particles have been randomly chosen 
in the intervals [0,0.35] and [$0^{\circ}, 45^{\circ}$], respectively.  

The test particles were integrated up to the first encounter within the Hill sphere 
of a giant planet, collision onto a planet or ejection. Those cut off conditions 
mean that plutinos moved from the stable zone of the resonance, so that they suffer an 
encounter with a planet, and then they do not belong any more to the resonant population.    
Therefore they are the escape conditions of plutinos.

\subsection{Selection of the long-term escapers}

\begin{figure}
\centering
\resizebox{\hsize}{!}{\includegraphics{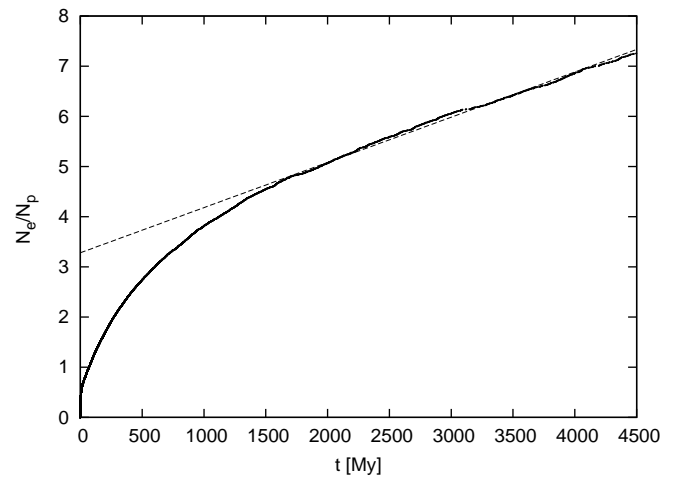}}
\caption{Cumulative number of escaped particles ($N_e$) with respect to the number of the 
surviving particles $N_{\bf p}$, versus the time $t$ in Myr. The fit to the plot 
for $t > 1.5$ Gyr is also showed.
	}
\label{tasaesc}
\end{figure}

From the $20,000$ initial particles, $17,577$ ($87.9 \%$) left out of the  integration 
at a given time either because of an encounter with a planet or an ejection. We have 
 21  particles that are ejected of the Solar System and
 $17,556$  that encounter Neptune or Uranus.  
The remaining $2,423$ particles ($12.1 \%$) keep inside the resonance up to the end 
of the integration. We will call the particles that leave the resonance, either because 
of an encounter or ejection, ``escaped plutinos''. 
 From our simulation we can calculate the rate of escape of particles from the 
resonance. In Fig. \ref{tasaesc} we plot the cumulative number of escape particles from the 
resonance ($N_e$) with respect to the number of the remaining particles $N_{ p}$, where 
$N_{p} = 20,000-N_e$, as a function of time. It can be seen that the number of escaped 
particles raise quickly at the beginning up to $t \sim 1.5$ Gyr. In this point, 
the slope of the curve changes and behaves roughly as a linear relation. This change 
of slope was already noticed by  Morbidelli (\cite{Morby97}) and is related with the 
time when the strongly chaotic region is completely depleted and the weakly chaotic 
region starts to be the dominant source of Neptune-encountering bodies. 
We fit to the plot for $t > 1.5$ Gyr, a linear relation given by:
\begin{equation}
N_e/N_{p} = a  t + b,
\label{nens}
\end{equation}
where: $a=9.00713 \times 10^{-10}  \pm 4.735 \times 10 ^{-13} yr^{-1} $ and 
 $b = 3.28079 \pm 0.001347$.

Morbidelli (\cite{Morby97}) sets out  that the slow diffusion region is the only active 
source of the 2:3 mean motion resonance that produce Neptune encountering bodies 
at present; then the slope of the linear relation, $a$, represents the present
 rate of escape of the particles from the 2:3 mean motion resonance. 
 We have noticed that there are very few particles of which the semimajor axis diverge 
from the resonant value before they  have an encounter with a planet or 
even if they never have an encounter with a planet. 
But this behaviour doesn't change the previous calculated rate of escape from the 
resonance.

\begin{figure}
\centering
\resizebox{\hsize}{!}{\includegraphics{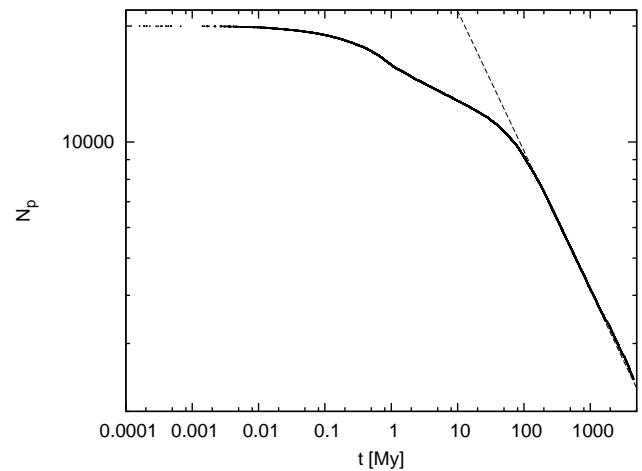}}
\caption{Number of remaining plutinos ($N_p$) versus the time $t$ in Myr. A power law 
fit to the plot for $t > 100$ Myrr is also showed.}
\label{npt}
\end{figure}

{\bf On the other hand, we plot in Fig. \ref{npt},  the number of remaining plutinos 
($N_p$) as a function of time. It can be seen a change of slope at $t \sim 100$ Myr, 
that was also already noticed by Morbidelli (\cite{Morby97}) and  Tiscareno \& Malhotra
 (\cite{Tiscareno09}). We can fit a power law to the number of surviving plutinos versus 
time for $t > 100$ Myr and is given by:

\begin{equation}
N_{p} =  k \,\,t\,^{\beta},
\label{nppot}
\end{equation}
where:  $k=7560950 \pm 34600 $ and 
 $\beta = -0.362672 \pm 0.0002372$.

Previous works have used power laws for fitting the number of surviving particles 
in the 2:3 mean motion resonance (Morbidelli \cite{Morby97}, 
Tiscareno \& Malhotra \cite{Tiscareno09}). In particular treating Morbidelli (\cite{Morby97}) data 
(see his Fig. 12) for $t>100$ Myr one can obtain $\beta = -0.32$. This is a value  
very close to our value of $\beta = -0.36$. As for the fitting of 
 Tiscareno \& Malhotra (\cite{Tiscareno09}), they obtained a somewhat steeper slope, 
although they fit the number of remaining particles vs time for the last 0.5 Gyr of 
their integration (this is from 500 Myrs to 1 Gyr). }

 The number of small objects in the plutino population is not well determined, 
since the present observational surveys can't cover all the small-sized objects. 
Then, the size distribution of plutinos is calculated from different surveys up 
to a given size, typically up to a radius $R \sim 30$ km. For objects with radius less 
than this size, the population could have a break (Kenyon et al. \cite{Kenyon08}, 
Bernstein et al. \cite{Bernstein04}, Elliot et al. \cite{Elliot05}).    
There are also theoretical models which account for accretion and fragmentation 
of planetesimals in the region of the Kuiper belt and also predict a broken power-law 
size distribution at a given radius (Kenyon et al.  \cite{Kenyon08}). 

de El\'{\i}a et al. (\cite{deelia08}), analyzed the size distribution of plutinos, taking 
into account the suggested mass of the population and the possible existence of a break in 
the distribution. Through their collisional evolution of plutinos they concluded that 
the existence of a break in the plutino size distribution should be a primordial feature. 
Then considering the analysis made by de El\'{\i}a et al. (\cite{deelia08}) and the three 
plutino size distributions proposed, we calculate the present number of plutinos with radius 
$R > 1$ km as $N_p \sim 10^8 - 10^9$, depending on the existence of the break.   
Then we will have a  present rate of escape of 1 to 10 plutinos with $R > 1$ km  
every 10 years.

We have 1183 particles that escape from the resonance after $t = 1.5$ Gyr and so 
they come from the slow diffusion region of the resonance. Since those particles 
would represent the present escaped plutinos, we identified their original orbital 
elements for our second integration. We will explain this in the 
following subsection.

\subsection{The Post-escape integration}

We numerically integrated again from $t = 0$ the 1183 particles that escape at 
$t > 1.5$ Gyr., in order now, to obtain their Post escape evolution. 
We integrated those particles with the same computing conditions than the 
previous integration. We used  
the EVORB code under the gravitational influence of the Sun and the four giant planets, 
the same step of integration of 0.5 years, but now we followed the integration for 10 Gyr. 
The particles were removed from the simulation when they either collide with a planet 
or the Sun,  or when they reach a semimajor axis greater than $1,000$ AU or they enter 
the region inside Jupiter's orbit ($r < 5.2$) where the perturbations of the 
terrestrial planets are not negligible. We recorded the orbital elements of the 
particles and the planets every $10^4$ years.  
In the following sections we are going to describe the results of this simulation.

\section{Escaped plutinos. General results }

The great majority of escaped plutinos have encounters with Neptune, so this planet 
is the main responsible for their post escape evolution. 
We register the encounters with the major planets at a distance of less than 3 Hill's 
radii. We have $1,945,643$  encounters with Neptune,  $534,557$ encounters with Uranus, 
$88,947$ encounters with Saturn and $1,309$ with Jupiter. However the encounters with 
Jupiter in particular are  reduced by the removal of objects at a distance of Jupiter.

From the 1183 initial particles, 1179  are removed from the integration and 4 particles 
remain in the integration. Those four particles have encounters with Uranus or/and Neptune, 
 they have short incursions to the Centaur zone and afterwards they are quickly transferred 
to a mean motion resonance in the SD remaining there up to the end of the integration. 
From the 1179 particles  removed from the integration  at some time,  790 ($67 \%$) 
are ejected, 385 ($32.7 \%$) reach the zone interior to Jupiter's orbit ($r < 5.2$), 
we have 4 ($0.3 \%$) collisions with the planets: 1 with Saturn, 1 with Uranus and 2 with 
Neptune. 
Those numbers can be compared with the ones obtained in the numerical simulation performed
by Di Sisto \& Brunini (\cite{Disisto07}); in particular the number of escaped plutinos 
that reach the orbit of Jupiter and the number of ejections are similar to those 
numbers for SDOs with low semimajor axis and perihelion distances less than 35 AU.   
 Also we can compare our results with the ones by Tiscareno \& Malhotra 
(\cite{Tiscareno09}). They obtained  that  $27 \%$ of escaped particles reach the zone 
of $r < 5.2$. This is a number slightly small than ours, but it could be due to the fact that 
Tiscareno \& Malhotra (\cite{Tiscareno09})'s result was obtained from the escaped particles 
from their 1 Gyr integration.

When a plutino escape from the resonance, it is transferred to the Scattered Disk (SD) 
zone ($q>30$ AU) or to the Centaur zone ($q<30$ AU).  This is also noticed by 
Tiscareno \& Malhotra (\cite{Tiscareno09}). The mean scale of time in reaching 
the Centaur zone is $670,000$ years and in reaching the SD zone is about 6 Myr.

\subsection{Distribution of orbital elements}

\begin{figure}
\centering
\resizebox{\hsize}{!}{\includegraphics{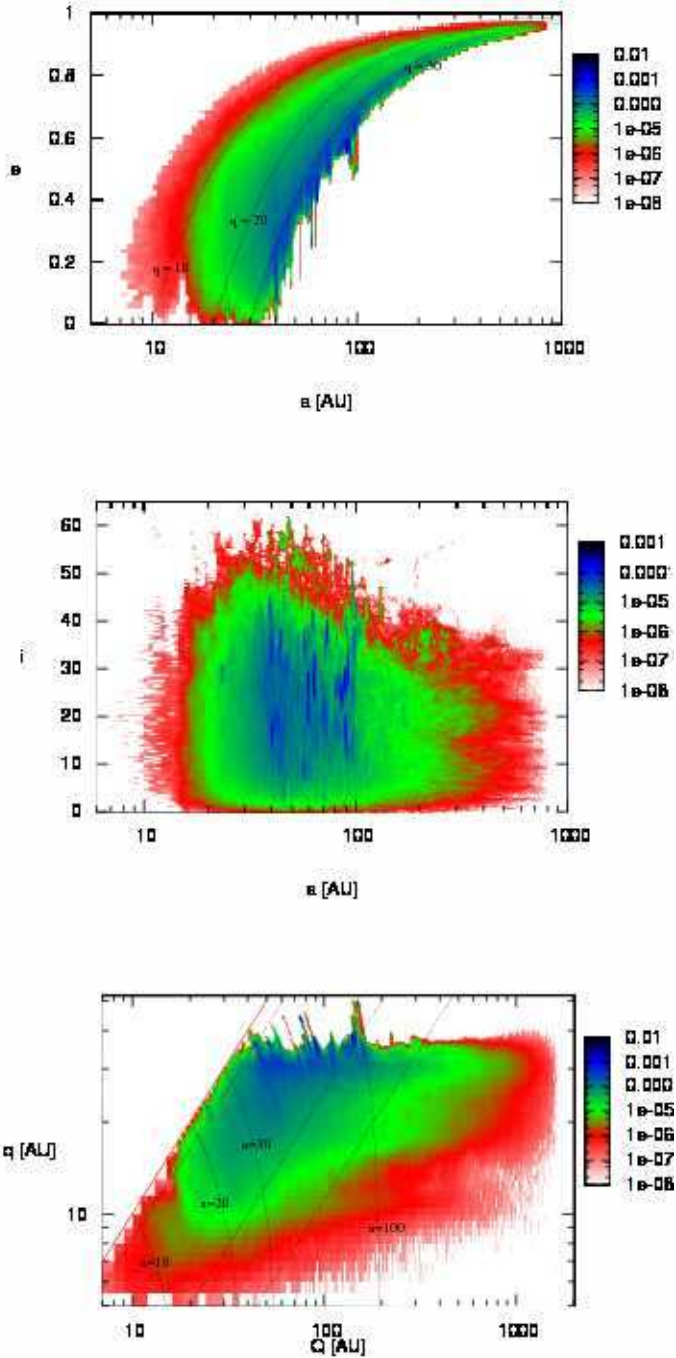}}
\caption{Time-weighed distribution of escaped plutinos in the orbital element 
space.
	}
\label{mapas}
\end{figure}

In Fig. (\ref{mapas}) we plot the time-weighed distribution of escaped 
plutinos in the orbital element space. There it is represented the probability 
distribution of finding an escaped plutino in the orbital element space. These 
plots assume time-invariability, so they don't represent the real case where 
plutinos are continuously leaving the resonance, passing through a certain  
zone out of the resonance and leaving the solar system, but they help to 
identify the densest and empties regions.   
As we can see, Neptune is the planet that mainly governs the post escape evolution
of escaped plutinos. This behavior can be seen in fig (\ref{mapas}) as the densest 
zone near Neptune's perihelion. 
 The densest zone in the orbital element space of escaped plutinos corresponds to the 
ranges of $30 < a < 100$ AU and $5^{\circ} < i < 40^{\circ}$. 
Also it can be seen several mean motion resonances densely populated, those corresponds to 
the blue lines denoted in the a vs i plot.

As we mentioned in the previous section, escaped plutinos take up the Centaur zone 
and the SDO zone. And in general they switch each of those populations during its 
dynamical evolution until it ends like one of them. This is obviously due to 
the presence of Neptune that lead the dynamical evolution of escaped plutinos. 
In particular, it is notable that  when escaped plutinos are transferred to the SD 
they are quickly locked into a mean motion resonance with Neptune. This behaviour 
is similar to the behavior of SDOs analyzed by Fern\'andez et al. (\cite{Fernandez04}) and 
Gallardo (\cite{Gallardo06}). The most densely populated mean motion resonances are: 
2:3N, 4:7N, 4:11N, 1:3N, 1:6U and 1:5N, where ``N'' means ``with Neptune'' and U 
means ``with Uranus''.  

The distribution of escaped plutinos in the Centaur zone is similar to 
that obtained from SDOs by Di Sisto \& Brunini (\cite{Disisto07}). We will analyze this 
contribution in the next section.

\section{Contribution to the Centaur population}

\subsection{Mean lifetime}

\begin{figure}
\centering
\resizebox{\hsize}{!}{\includegraphics{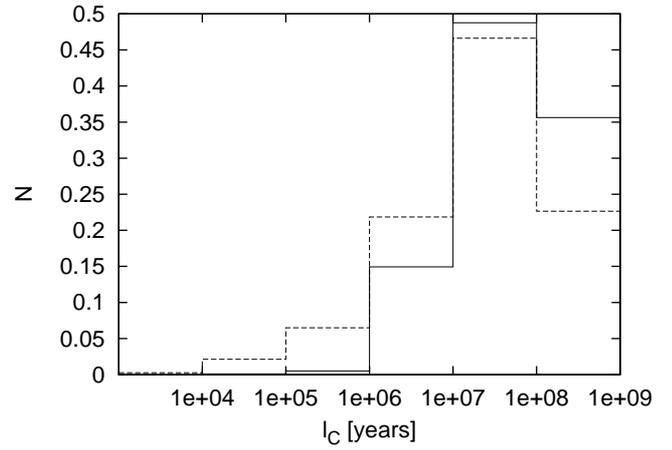}}
\caption{Normalized fraction of escaped plutinos against lifetime in the Centaur zone 
(filled line) and normalized fraction of SDOs that enter the Centaur zone against 
lifetime there (dashed line).
	}
\label{ndet}
\end{figure}

From the 1183 particles that escape from plutinos  1179 particles enter to the 
Centaur zone. The four remaining escaped plutinos are ejected by the first encounter 
with a planet (Neptune in all these cases) within 1 Hill's radii.
The escaped plutinos have a mean lifetime in the Centaur zone of 
$ l_C = 108$ Myr. This is greater than the mean lifetime of Centaurs from the SD of 
$72$ Myr (Di Sisto \& Brunini \cite{Disisto07}).  
 In Fig. (\ref{ndet}) we plot the normalized fraction of escaped plutinos against intervals 
of lifetime in the Centaur zone. Also it is plotted, for a comparison, the distribution 
of lifetimes in the Centaur zone for objects from the SD. As we can see Centaurs from 
plutinos have greater lifetime than Centaurs from SDOs, and the great majority of 
lifetimes of plutinos in the Centaur zone is greater than  1 Myr.  So escaped plutinos 
spend long time in the Centaur zone, from 1 Myr to 1000 Myr. From our numerical 
simulation, we noticed that the mean lifetime of plutinos in the Centaur zone,  
as a function of the initial inclination, has nearly 
the same behavior than Centaurs from SDOs (see Fig.5 in Di Sisto \& Brunini \cite{Disisto07}),
 but they show higher values in each bin of initial inclination. So the difference 
in the mean lifetime of plutino-Centaurs and SDO-Centaurs is not dependent on the 
initial inclinations.  

\begin{figure}
\centering
\resizebox{\hsize}{!}{\includegraphics{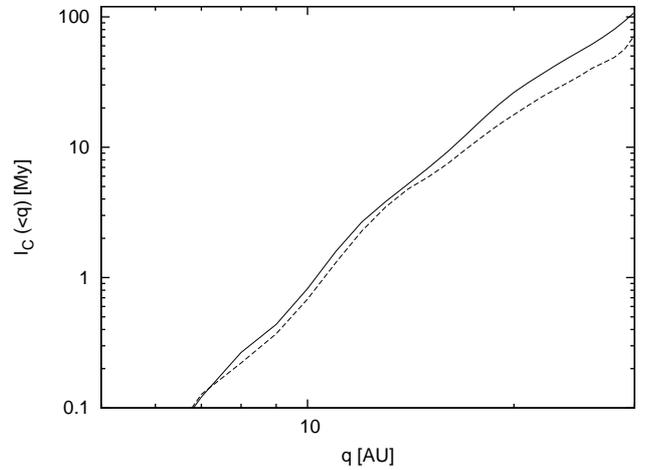}}
\caption{Mean lifetime of Plutinos (filled line) and of SDOs (dashed line) in the Centaur 
zone with perihelion distances less than a given value.
	}
\label{tdeq}
\end{figure}

In Fig (\ref{tdeq}) we plot the lifetime of plutinos in the Centaur zone 
with perihelion distances $q$ less than a given value and also the mean lifetime of 
SDOs in the Centaur zone, as a comparison.  For building this plot, we count for each 
particle ($i$) the time that it spends with $q$ less than a given value $q_0$, 
($dt_i (q<q_0)$). Also we count the number $N$ of particles that remain for a certain time 
with $q < q_0$. Then the mean lifetime of the particles with $q<q_0$ would be 
$l_C (<q_0) = \frac{\sum_{i=1}^N dt_i (q<q_0)}{N}$.

\begin{figure}
\centering
\resizebox{\hsize}{!}{\includegraphics{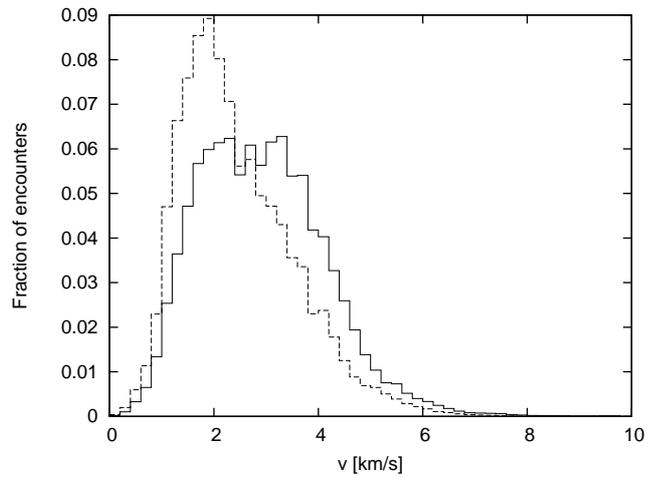}}
\caption{Normalized distribution of encounter velocities with the major planets 
of SDOs (dashed line) and escaped plutinos (solid line).
	}
\label{venc}
\end{figure}

We can see a stronger dependence of 
Centaur lifetime on their perihelion distance for plutino-Centaurs than for 
SDO-Centaurs. However the greatest differences are found for large perihelion 
distances, and for perihelion distances less than $\sim 15$ AU both populations
have similar lifetimes.  
Then, escaped-plutinos live more time than SDOs in the greater-perihelion Centaur 
zone, causing a slower diffusion to the inner Solar System of escaped-plutino
 orbits than of SDOs orbits.  We suggest that this fact is due to the difference in the 
relative encounter velocities with the major planets of SDOs and plutinos. In Fig. 
(\ref{venc}) we show the distribution of the relative encounter velocities of 
escaped plutinos (solid line) and SDOs (dotted line) with the four giant planets. 
It is clearly seen that SDOs have smaller encounter velocities than plutinos. 
The mode of the encounter velocities is equal to $3.22$ km/s for plutinos and 
$1.86$ km/s for SDOs, and the mean values are  equal to $2.92$ km/s for plutinos 
and $2.45$ km/s for SDOs. Orbits with smaller relative encounter velocities suffer bigger 
changes in velocity, and then also in the orbital energy. Successive encounters with 
the major planets make the objects to evolved to the inner Solar System or to the 
transneptunian region up to ejection. So bigger changes in the orbits lead to accelerate 
those transitions.  We conclude that's why SDOs have a mean lifetime smaller than 
that of escaped plutinos.

\subsection{Number of plutino-Centaurs}

\begin{figure}
\centering
\resizebox{\hsize}{!}{\includegraphics{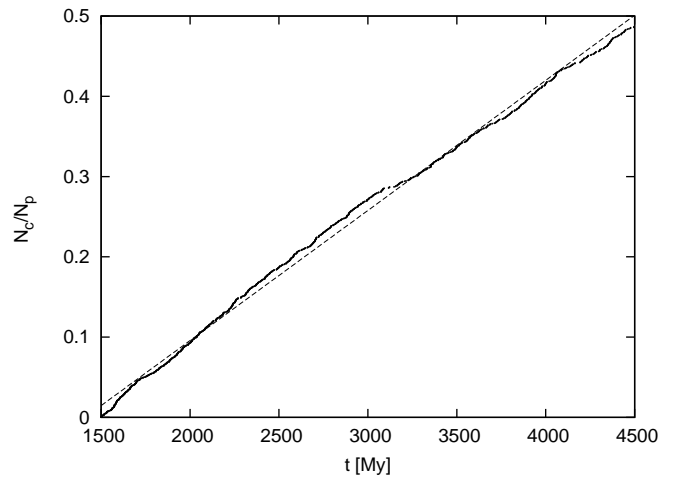}}
\caption{Cumulative number of escaped plutinos injected into the Centaur zone 
($N_c$) with respect to the number of the remaining plutinos $N_p$, as a function of time. 
The dashed line is the fitting to the data.
	}
\label{tasaicent}
\end{figure}

In order to calculate the number of escaped plutinos located at present in the Centaur
 population, we calculate the present rate of injection of escaped plutinos into the 
Centaur zone. As we have mentioned we consider that the long-term escapers of the plutino 
population, i.e., those that escape after $t = 1.5$ Gyr, represent the present plutino
espapers.  In Fig, (\ref{tasaicent}) we plot the cumulative number of escaped plutinos 
injected into the Centaur zone ($N_c$) with respect to the number of the remaining plutinos 
($N_p$), as a function of time. We have taken into account the fact that a plutino escape 
from the population when it has an encounter with a planet or it is ejected, so the number 
of the remaining plutinos depends on plutinos injected to the Centaur zone  but also 
on the escaped plutinos that are permanently or transitory injected into the SD zone. 
 As Fig. (\ref{tasaicent}) shows, the ratio $N_c/N_p$ is well fitted by a linear relation, 
given by: 
   
\begin{equation}
N_c/N_p = c \, t + d,
\label{ncnp}
\end{equation}

\noindent
where: $c=1.62076 \times 10^{-10}  \pm 8.502 \times 10 ^{-14}  yr^{-1}$  and 
 $d = -0.228442 \pm 0.000242$.

The slope of the linear relation, $c$, represents the present rate of injection of 
plutinos from the 2:3 mean motion resonance into the Centaur zone. 

The present number of escaped plutinos in the Centaur population  can be calculated 
by differentiating Eq. 2: $dN_C / dt = c N_p  $ where we have assumed that $N_p$ 
is constant during an interval of time $dt$. Then, in order to calculate the present 
number of plutino-Centaurs we  take $dt = l_C$ and now assume that 
$N_p$ is constant during the lifetime of plutino-Centaurs.

\begin{equation}
N_c = c \, N_p \,  l_C 
\label{np}
\end{equation}

\noindent
where $l_C$  is the mean lifetime in the Centaur zone.

 Taking from de El\'{\i}a et al. (\cite{deelia08}) the present number of plutinos with 
radius $R > 1$ km as $N_p \sim 10^8 - 10^9$;  the rate of injection of plutinos with 
radius greater than 1 km to the Centaur zone will be between 
1.6 to 16 plutinos every 100 years. This is between 250 and 25 times less than the 
rate of injection of SDOs to the Centaur zone obtained by Di Sisto \& Brunini 
(\cite{Disisto07}). Then the present number of plutino-Centaurs with radius greater of 
1 km would be between $1.8 \times 10^{6} - 1.8 \times 10^{7}$. 

Di Sisto \& Brunini (\cite{Disisto07}) estimated a number of Centaurs with $R>1$ km coming 
from the SD of $\sim 2.8 \times 10^8$, then Centaurs coming from plutinos would represent 
a fraction of less than  $6 \%$  of the total Centaur population. That is to say that 
the plutino population is a secondary source of Centaurs, comparable to the 
contribution of the low eccentricity transneptunian objects according to the 
estimations of Levison \& Duncan (\cite{Levison97}) of $1.2 \times 10^7$.

\subsection{Orbital evolution of escaped plutinos in the Centaur zone}

Di Sisto \& Brunini (\cite{Disisto07}) analyzed the dynamical behavior of the SDOs that enter 
the Centaur zone, and found four classes of dynamical evolution. The Centaurs, then, 
behaves as one of the classes or as a combination of them. For the reason of completeness, 
we will briefly describe the four classes here. 
The first type is characterized  by the conservation of the perihelion
distance in a range of values between that of  Saturn and Neptune's orbit, 
the conservation or the very slow variation of the  perihelion  longitude, 
and eccentricities  greater  than $\sim 0.8$. The
characteristics  of   this  evolution  make   the  orbit  go   into  a
``pseudo-stable'' state  during which the encounters with the planets are
avoided  or  very  weak, causing  a  very  slow  variation of the orbit
orientation. 
The second type of objects are those that show ``resonance hopping'',
(this  is  a  phenomenon  in  which  objects  move  quickly  from  one
resonance,  in this  case with  Neptune, to  another) combined  with a
behavior similar to the first  one but with less constant eccentricity
values  and constant  perihelion  distances for  shorter intervals  of
time (Tiscareno \& Malhotra, \cite{Tiscareno03}). 
This type of objects  have also transfers between  mean motion resonances and
 Kozai resonances.  A Kozai  resonance occurs when the argument of
pericentre, $w$, librates about a constant value. For low inclinations
it is possible for $w$ to librate  about $w = 0 ^{\circ}$ and $w = 180
^{\circ}$, and for large inclinations about $w = 90 ^{\circ}$ and $w =
270 ^{\circ}$.  The semimajor axis  of the object remains constant but
the eccentricity and the inclination  of the orbit are coupled in such
a  way  that  $e$ is  a  maximum  when  $i$  is  a minimum,  and  vice
versa.
In  these two  first types,  the objects  have casual  encounters with
 Neptune  and Uranus,  sometimes also  with Saturn,  but they  are not
 strong  enough  to drastically  change  their  orbit causing a kind of 
stable orbit in the Centaur zone.
In the third  type of objects, we group those  that have the behaviors
of the first and second types, but they have perihelion distances near
Neptune. So,  the objects are continuously  entering and leaving the
Centaur zone.  
The last type of objects are  those that enter a mean motion resonance
or Kozai resonance for almost all their lifetime as a Centaur.

Bailey \& Malhotra (\cite{Bailey09}), have analyzed the chaotic behavior of the 
known Centaurs. Their analysis has revealed that two types of chaotic evolution are 
quantitatively distinguishable. One random walk-type behavior and an orbital 
evolution dominated by intermittent resonance sticking. 
These two dynamical classes embrace the four ones already found by Di Sisto \& Brunini 
(\cite{Disisto07}).
Bailey \& Malhotra (\cite{Bailey09}) also found that these two types of behavior are 
correlated with Centaur dynamical lifetime.

\begin{figure}
\centering
\resizebox{\hsize}{!}{\includegraphics{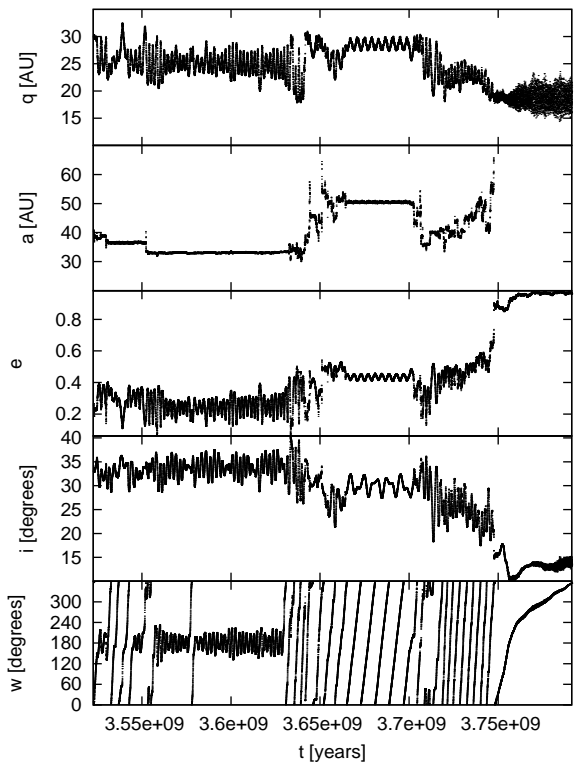}}
\caption{Dynamical Evolution of the orbital elements of an escaped plutino injected 
into the Centaur zone.
	}
\label{evolcent}
\end{figure}

In this paper we have analyzed the orbital evolution of escaped plutinos in the Centaur 
zone, and we have found that they can be grouped into the four dynamical classes proposed 
by Di Sisto \& Brunini (\cite{Disisto07}). There are more particles that have the dynamical 
behaviors of the 
second class. It is notable the great frequency of the presence of Kozai resonances in 
all the four classes. In the Centaur zone, mean motion resonances and Kozai resonances 
are more frequent for semimajor axis between 30 AU and 50 AU. 
For example in Fig. (\ref{evolcent}) we show the dynamical evolution of one plutino-Centaur  
of the second class that shows transfers between  mean motion resonances and
 Kozai resonances great part of the time. 
There are also some particles that once they escape, they return to a plutino state 
for some time and then continue their evolutions until their final state. 
Also there  are particles that  exhibit a ``hand off'' from the gravitational
control of one planet to another until crossing the Jupiter's orbit. The time scale 
of this hand off is whether as short as few Myr or as long as some hundreds of Myr. 
The particles that have the shorter lifetimes (say some Myr) carry
out short incursions to  the Centaur zone, or a handing down from a Jovian planet 
to the next inside until Jupiter, eventually passing through mean motion resonances, 
or a quick diffusion through all the inter giant-planetary zone until ejection or 
injection to Jupiter orbit.

\section{Conclusions}

We have performed two numerical simulations in order to first obtain particles 
representative to the plutinos that are escaping at present from the resonance and 
second in order to describe their dynamical post escape evolution.    
In the first simulation we integrate $20,000$ initial particles in the 2:3 resonance 
and find that $\sim 88 \%$ of the particles left out the integration and the rest remain
in it. Considering a plutino population with radius grater than 1 km 
of $N_p \sim 10^8 - 10^9$, we obtained a present rate of escape of plutinos 
between 1 and 10 every 10 years.  
From this integration we selected those particles that are representatives of the 
present escape plutinos and performed a second integration. From this last integration 
we obtained the dynamical evolution of plutinos once they escape from the resonance. 
From the 1183 initial particles, 1179  were removed from the integration and 
4 remain in it. 
From the 1179  removed particles, 787 are ejected, 385 reached the Jupiter's zone 
and 4 collide with the planets.  

We found that the great majority of escaped plutinos have encounters with Neptune, and 
this planet governs their dynamical evolution. 
When a plutino escape from the resonance, it is transferred to the SD zone ($q>30$ AU) 
or to the Centaur zone ($q<30$ AU) but it eventually switches to those population, due 
to the dynamical influence of Neptune. 
   
The densest zone in the  orbital element space of escaped plutinos corresponds to the ranges 
$30 < a < 100$ AU and $5^{\circ} < i < 40^{\circ}$ and perihelions near the orbit of Neptune.  
 When escaped plutinos are transferred to the SD they are quickly locked into 
a mean motion resonance with Neptune (similar to the behavior of SDOs analyzed by 
Fern\'andez et al. (\cite{Fernandez04}) and Gallardo (\cite{Gallardo06}).
In the Centaur zone (this is the zone of $q < 30$ AU ) the distribution of escaped 
plutinos is similar to that of SDOs in the Centaur zone obtained by Di Sisto \& 
Brunini (\cite{Disisto07}). 

The orbital evolution of escaped plutinos in the Centaur zone can be grouped into 
the four dynamical classes proposed by Di Sisto \& Brunini  (\cite{Disisto07}). 
There are more particles that have the dynamical behavior of the 
second class and it is notable the great frequency of the presence of Kozai resonances in 
all the four classes. There are also several mean motion resonances densely populated in the 
ranges of $30 < a < 50$ AU.  

  The escaped plutinos have a mean lifetime in the Centaur zone of 
$  108$ Myr, greater than that of Centaurs from SD of $  72$ Myr . 
 Escaped-plutinos live more time than SDOs in the greater-perihelion Centaur 
zone, causing a slower diffusion to the inner Solar System of escaped-plutino
 orbits than of SDOs orbits. 

The present rate of injection of plutinos with radius greater than 1 km to the Centaur 
zone is between 1.6 to 16 plutinos every 100 years and the number 
of plutino-Centaurs with radius greater than 1 km would be between 
$1.8 \times 10^{6} - 1.8 \times 10^{7}$. Both, the rate of injection and the number of 
Centaurs from plutinos are much less than the contribution from the SD obtained 
by Di Sisto \& Brunini  (\cite{Disisto07}). Then, plutinos would represent a secondary source 
of Centaurs and their contribution would be a fraction of less than  $6 \%$ 
 of the total Centaur population. 

\vspace*{1cm}

\noindent{\bf Acknowledgments:} We thank  Matthew S. Tiscareno who, as referee, 
made valuable comments that helped to improve this manuscript.

\end{document}